\documentclass[prb,reprint,superscriptaddress,citeautoscript]{revtex4-1}

\usepackage[applemac]{inputenc}
\usepackage[T1]{fontenc}
\usepackage[american]{babel}
\usepackage[scaled]{helvet}
\usepackage{graphicx,bm,textcomp,courier,multirow,hyperref,color,pdfpages,subfigure}
\usepackage{amsmath,amssymb}
\usepackage[normalem]{ulem}	

\newcommand{\Dirac}[3]{\left\langle #1 \left| #2\right| #3\right\rangle}

\newcommand{\op}[2]{\left.{\left| #1\right\rangle\left\langle #2\right|}\right.}

\newcommand{\ket}[1]{\left|\left. #1 \right\rangle\right.}

\newcommand{\abs}[1]{\left| #1 \right|}

\newcommand{\pos}[1]{^{\left(#1\right)}}
\newcommand{\figureshortname}{Fig.}
\newcommand{\equationshortname}{Eq.}
\newcommand{\tableshortname}{Tab.}

\newcommand{\eref}[1]{\equationshortname~\eqref{#1}}
\newcommand{\sref}[1]{Sec.~\ref{#1}}
\newcommand{\aref}[1]{Appx.~\ref{#1}}
\newcommand{\cref}[1]{Chapter~\ref{#1}}
\newcommand{\fref}[1]{\figureshortname~\ref{#1}}
\newcommand{\tref}[1]{\tableshortname~\ref{#1}}
\newcommand{\rcite}[1]{Ref.~[\onlinecite{#1}]}
\newcommand{\mrcite}[2]{Refs.~[\onlinecite{#1},\onlinecite{#2}]}
\newcommand{\mmrcite}[3]{Refs.~[\onlinecite{#1},\onlinecite{#2},\onlinecite{#3}]}

\graphicspath{{Pictures/}}

\begin{document}
\def\sectionautorefname{Sec.}

\title{Inverted Singlet-Triplet Qubit Coded on a Two-Electron Double Quantum Dot}

\author{Sebastian Mehl}
\email{s.mehl@fz-juelich.de}
\affiliation{Peter Grünberg Institute (PGI-2), Forschungszentrum Jülich, D-52425 Jülich, Germany}
\affiliation{JARA-Institute for Quantum Information, RWTH Aachen University, D-52056 Aachen, Germany}
\author{David P. DiVincenzo}
\affiliation{Peter Grünberg Institute (PGI-2), Forschungszentrum Jülich, D-52425 Jülich, Germany}
\affiliation{JARA-Institute for Quantum Information, RWTH Aachen University, D-52056 Aachen, Germany}

\date{\today}

\begin{abstract}
The $s_z=0$ spin configuration of two electrons confined at a double quantum dot (DQD) encodes the singlet-triplet qubit (STQ). We introduce the inverted STQ (ISTQ) that emerges from the setup of two quantum dots (QDs) differing significantly in size and out-of-plane magnetic fields. The strongly confined QD has a two-electron singlet ground state, but the weakly confined QD has a two-electron triplet ground state in the $s_z=0$ subspace. Spin-orbit interactions act nontrivially on the $s_z=0$ subspace and provide universal control of the ISTQ together with electrostatic manipulations of the charge configuration. GaAs and InAs DQDs can be operated as ISTQs under realistic noise conditions.
\end{abstract}

\maketitle

\section{Introduction}
Encoded spin qubits in a two-electron configuration have become popular since the seminal experiment by Petta et al. \cite{petta2005} Single electrons are trapped using gate-defined quantum dots (QDs) in semiconducting nanostructures \cite{hanson2007-2}. The spin is used as the information carrier \cite{loss1998}. We consider the qubit encoding using the $s_z=0$ spin subspace of two electrons \cite{levy2002,taylor2005,hanson2007-1}. The passage between different charge configurations realizes single-qubit control electrostatically. Applying voltages at metallic gates close to the structure enables the  transfer of electrons between the QDs. The $\left(1,1\right)$ configuration labels separated electrons on the two QDs; two electrons occupy a single QD in $\left(2,0\right)$ and $\left(0,2\right)$.

In this paper, we explore a two-electron double quantum dot (DQD) under the influence of magnetic fields and spin-orbit interactions (SOIs). The qubit is encoded in the $s_z=0$ subspace of two electrons using the singlet $\ket{S}$ and spinless triplet $\ket{T}$ states, similarly to common singlet-triplet qubits (STQs) \cite{levy2002,taylor2005,hanson2007-1}. Our setup has an energy degeneracy of $\ket{S}$ and $\ket{T}$ in $\left(1,1\right)$ that is a consequence of the competition between the confining potential and the Coulomb interactions. In the absence of SOIs, the orbital contributions from the out-of-plane magnetic fields favor triplets, while the confining potential favors singlets. We call this qubit inverted STQ (ISTQ) because it differs from normal STQs by the occurrence of a singlet-triplet inversion. We realize an ISTQs with one strongly confined QD and one weakly confined QD. $\ket{T}$ is the ground state in $s_z=0$ for one QD when it is doubly occupied, but the other QD has a singlet ground state. SOIs couple $\ket{S}$ and $\ket{T}$. In contrast to the setup with two QDs differing significantly in size, it was argued that SOIs act trivially on the $s_z=0$ subspace for two identical QDs \cite{baruffa2010-1,baruffa2010-2}.

The encoding in the $s_z=0$ subspace is optimal because the qubit encoding is protected from hyperfine interactions. Nuclear spins generate local magnetic field fluctuations $\bm{B}_{\text{hyp}}$. Mainly the component $B_{\text{hyp}}^\shortparallel$ parallel to the external magnetic field $\bm{B}$ influences the $s_z=0$ subspace \cite{coish2005}. Fluctuations in $B_{\text{hyp}}^\shortparallel$ are low frequency and can be corrected using refocusing techniques \cite{bluhm2011,neder2011}. In particular, the ISTQ is superior to the two-electron encoding that uses the singlet state $\ket{S}$ and the $s_z=1$ triplet state $\ket{T^+}$ \cite{petta2010,ribeiro2010,ribeiro2013-1,ribeiro2013-2}. There is also an energy degeneracy of $\ket{S}$ and $\ket{T^+}$ in this setup, but hyperfine interactions induce noise with larger weights at higher frequencies \cite{neder2011}.

The main purpose of this paper is to explore the ISTQ encoding. We show that SOIs act nontrivially on the $s_z=0$ subspace. The influence of SOIs can be described by an effective magnetic-field difference between the QDs. The effective local magnetic field depends on the confining potential of the wave functions. ISTQs are controlled using electrostatic voltages, which tune the DQD between different charge configurations. DQDs that consist of QDs with different sizes realize ISTQs that can be operated in the presence of realistic noise sources. A DQD that is coded using two distinct QDs gives also other perspectives: A strongly confined QD is favorable for the initialization and the readout of STQs. A weakly confined QD may be favorable for qubit manipulations \cite{mehl2013}. We are convinced that this setup is likely to be explored as the search for alternative spin qubit designs continues \cite{shi2014,kim2014,higginbotham2014}. Operating STQs coded using two QDs with different sizes as ISTQs is achieved by applying sufficiently large out-of-plane magnetic fields.

The organization of this paper is as follows. \sref{C6-sec:Model} introduces the model to construct ISTQs and describes the qubit encoding. \sref{C6-sec:SOI} characterizes SOIs as a source to influence the $s_z=0$ subspace. We describe different possibilities to manipulate the ISTQ in \sref{C6-sec:Manipulate} and discuss its performance in \sref{C6-sec:Discussion}.

\section{Model
\label{C6-sec:Model}}
Our study includes the orbital Hamiltonian $\mathcal{H}_0$, external magnetic fields $\mathcal{H}_1$, and SOIs $\mathcal{H}_2$. The orbital Hamiltonian for two electrons in gate-defined lateral DQDs is described by:
\begin{align}
	\mathcal{H}_{0}=\sum_{i=1,2}\left[\frac{\bm{\wp}^2_i}{2m}+V\left(\bm{x}_i\right)\right]+V\left(\bm{x}_1,\bm{x}_2\right).
	\label{C6-eq:H0}
\end{align}
The orbital contributions of the magnetic-field component perpendicular to the lateral direction (called the z-direction) are included by the kinematic momentum operator $\bm{\wp}=\frac{\hbar}{i}\bm{\nabla}+e\bm{A}$. $e>0$ is the electric charge, $m$ is the effective mass, and $\bm{A}=\frac{B_z}{2}\left(-y,x,0\right)^T$ describes orbital effects from the out-of-plane magnetic-field component $B_z$ in the symmetric gauge. Orbital contributions from in-plane magnetic fields are weak for strong confining potentials in the z-direction. $V\left(\bm{x}\right)$ is the single-particle potential that includes external electric fields. Two QDs are present at the positions $\left(\pm a,0,0\right)^T$. $V\left(\bm{x}_1,\bm{x}_2\right)$ is the Coulomb interaction. Magnetic fields couple directly to the spins through the Zeeman Hamiltonian:
\begin{align}
	\label{C6-eq:Zeem}
	\mathcal{H}_{1}=\frac{g\mu_B}{2}\bm{B}\cdot\sum_{i=1,2}\bm{\sigma}_i.
\end{align}
$\bm{\sigma}=\left(\sigma_x,\sigma_y,\sigma_z\right)^T$ is the vector of Pauli matrices, $\bm{B}$ is the magnetic field, $g$ is the g-factor, and $\mu_B$ is the Bohr magneton.

We include two orbitals at each QD: the single-dot ground $\big\{ \ket{L} , \ket{R} \big\}$ and the single-dot excited states $\big\{ \ket{\overline{L}}, \ket{\overline{R}} \big\}$. We consider only the $s_z=0$ subspace, since states with $s_z\ne 0$ are far away in energy during qubit manipulations. The wave functions of the singlet state (S) and spinless triplet state (T) of different charge configurations $\left(n_L,n_R\right)$ are
\begin{align}
\label{C6-eq:states_begin}
\ket{S_{1,1}}		&	=\frac{1}{\sqrt{2}}\left(
c_{L,\uparrow}^\dagger c_{R,\downarrow}^\dagger-
c_{L,\downarrow}^\dagger c_{R,\uparrow}^\dagger
\right)\ket{0},\\
\ket{S_{2,0/0,2}}	&	=\left(
c_{L/R,\uparrow}^\dagger c_{L/R,\downarrow}^\dagger
\right)\ket{0},\\
\label{C6-eq:state_triplet}
\ket{T_{1,1}}		&	=\frac{1}{\sqrt{2}}\left(
c_{L,\uparrow}^\dagger c_{R,\downarrow}^\dagger+
c_{L,\downarrow}^\dagger c_{R,\uparrow}^\dagger
\right)\ket{0},\\
\ket{T_{2,0/0,2}}	&	=\frac{1}{\sqrt{2}}\left(
c_{L/R,\uparrow}^\dagger c_{\overline{L}/\overline{R},\downarrow}^\dagger+
c_{L/R,\downarrow}^\dagger c_{\overline{L}/\overline{R},\uparrow}^\dagger
\right)\ket{0},
\label{C6-eq:states_end}
\end{align}
where $\ket{0}$ is the vacuum state and $c^\dagger_{i\sigma}$ is the creation operator of an electron in orbital $i$ with spin $\sigma$. We use a Hubbard model to describe the $\left(2,0\right)$, $\left(1,1\right)$, and $\left(0,2\right)$ configurations \cite{burkard1999,hu2000}. The electrons are on separate QDs in $\left(1,1\right)$. The orbital ground states are filled with two electrons for the singlets $\ket{S_{2,0}}$ and $\ket{S_{0,2}}$; the Pauli exclusion principle requires that electrons fill different orbitals for $\ket{T_{2,0}}$ and $\ket{T_{0,2}}$. Orbital effects of $\mathcal{H}_0$ and $\mathcal{H}_1$ are described by
\begin{align}
\nonumber
\mathcal{H}=&
\left(
\begin{array}{ccc}
0 & t_s^L & t_s^R\\
t_s^L & U_L+\epsilon & 0\\
t_s^R & 0 & U_R+\Omega_{\left(0,2\right)}-\epsilon
\end{array}
\right)\\&
\oplus
\left(
\begin{array}{ccc}
0 & t_t^L & t_t^R\\
t_t^L & U_L+\Omega_{\left(2,0\right)}+\epsilon & 0\\
t_t^R & 0 & U_R-\epsilon
\end{array}
\right)
\label{C6-eq:Ham}.
\end{align}
\eref{C6-eq:Ham} is written in the basis $\left\{\ket{S_{1,1}},\ket{S_{2,0}},\ket{S_{0,2}},\ket{T_{1,1}},\ket{T_{2,0}},\ket{T_{0,2}}\right\}$. The real constants $t_{s,t}^{L,R}$ characterize the spin-conserving hopping processes of electrons from $\left(1,1\right)$ towards two electrons on the same QD. The relative energies of $\left(2,0\right)$, $\left(1,1\right)$, and $\left(0,2\right)$ are tunable by voltages at gates near the left and right QD; we model their influence as a modification of the addition energies $U_{L}\rightarrow U_{L}+ \epsilon$ and $U_{R}\rightarrow U_{R}- \epsilon$. The left QD is doubly occupied for $\epsilon\rightarrow-\infty$ (and, similarly, the right QD for $\epsilon\rightarrow\infty$). The electrons are separated on different QDs for $\epsilon\sim 0$.

\begin{figure}[htb]
\centering
\includegraphics[width=0.49\textwidth]{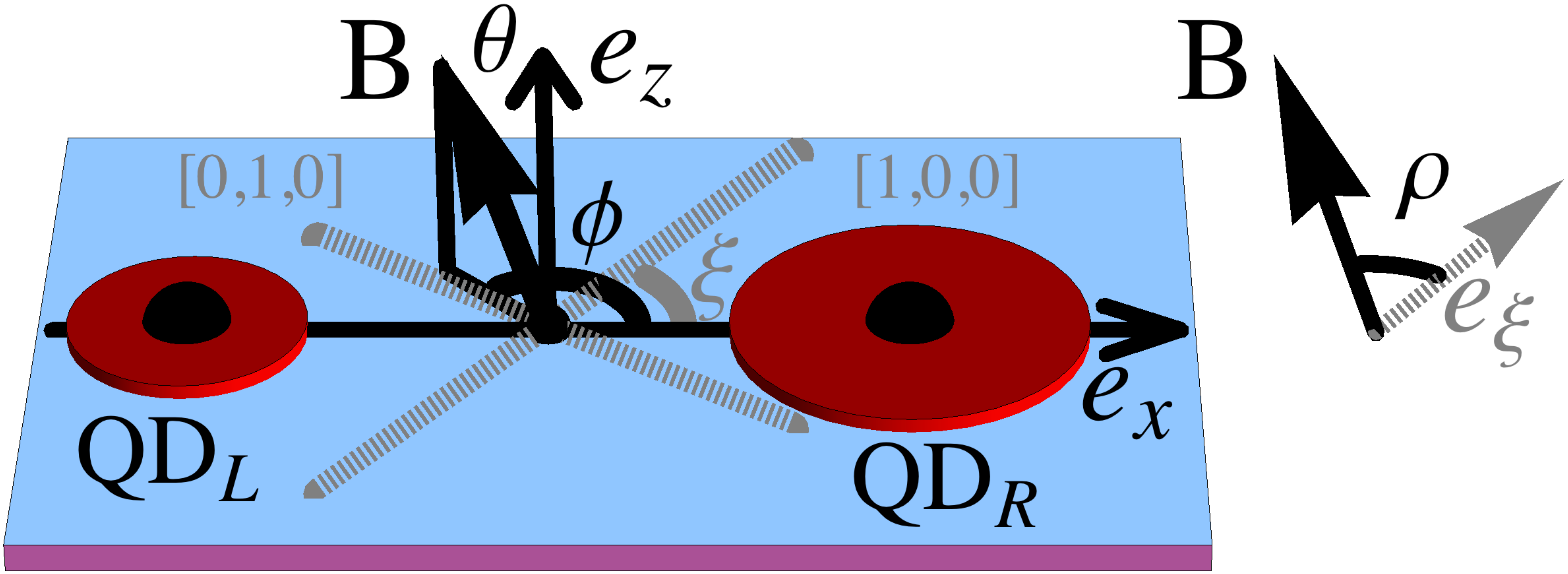}
\caption{STQ coded on an asymmetric DQD. Each QD contains one electron. Modifications of the confining potentials allow an electron transfer to reach a doubly occupied QD. $\text{QD}_L$ has a two-electron singlet ground state; for $\text{QD}_R$ the spinless triplet state is lower in energy. The out-of-plane component $B_z$ of the magnetic field favors triplets, but the confining energy favors singlets. The magnetic field $\bm{B}$ is tilted by the angle $\phi$ from the dot-connection axis $\bm{e}_x$ and by the out-of-plane angle $\theta$ from $\bm{e}_z$. The $\left[1,0,0\right]$-direction of the lattice is rotated by the angle $\xi$ from $\bm{e}_x$. We introduce additionally the rotation angle $\rho$ between $\bm{B}$ and the $\left[1,0,0\right]$-direction.
\label{C6-fig:1}}
\end{figure}

As above, one needs to overcome the charging energies $U_{L}$ of the left QD or $U_{R}$ of the right QD to add two electrons to the same QD. One QD (e.g., $\text{QD}_L$) is in the normal configuration and has a singlet ground state, but $\ket{T_{0,2}}$ is the ground state of $\text{QD}_R$. The singlet is the ground state in the absence of magnetic fields \cite{lieb1962}. Doubly occupied QDs with $E_{T}<E_{S}$ are obtained at finite out-of-plane magnetic fields also for $s_z=0$ \cite{merkt1991,wagner1992}. Finite values of $B_z$ decrease the sizes of the orbital wave functions and raise the Coulomb repulsions between the electrons. Electrons prefer to minimize the Coulomb repulsion, which makes triplets favorable. The inversion from a singlet to a triplet ground state was experimentally detected at $B_z=1.5~\text{T}$ in elongated GaAs QDs \cite{zumbuhl2004}. A theoretical study predicts an orbital singlet-triplet inversion at $B_z=0.5~\text{T}$ in weakly confined, circular GaAs QDs \cite{baruffa2010-2}. However, ISTQs only require a triplet ground state for one of the two QDs, which is realized for one strongly confined QD and one weakly confined QD (cf. \fref{C6-fig:1}).\cite{[{The singlet-triplet inversion of gate-defined, asymmetric DQDs was proven in numerical simulations of GaAs heterostructures, }]hiltunen2014} $U_{L}+\Omega_{\left(2,0\right)}$ and $U_{R}+\Omega_{\left(0,2\right)}$ are the energies to reach the first excited, doubly occupied states,
\begin{align}
\Omega_{\left(2,0\right)}&=E_{T_{2,0}}-E_{S_{2,0}}>0,\ \ \ 
\Omega_{\left(0,2\right)}=E_{S_{0,2}}-E_{T_{0,2}}>0,
\end{align}
are the energy differences of the doubly occupied states. We neglect matrix elements between $\left(2,0\right)$ and $\left(0,2\right)$ of the same spin \cite{burkard1999} because their contributions are weak.

\begin{figure}[htb]
\centering
\includegraphics[width=0.49\textwidth]{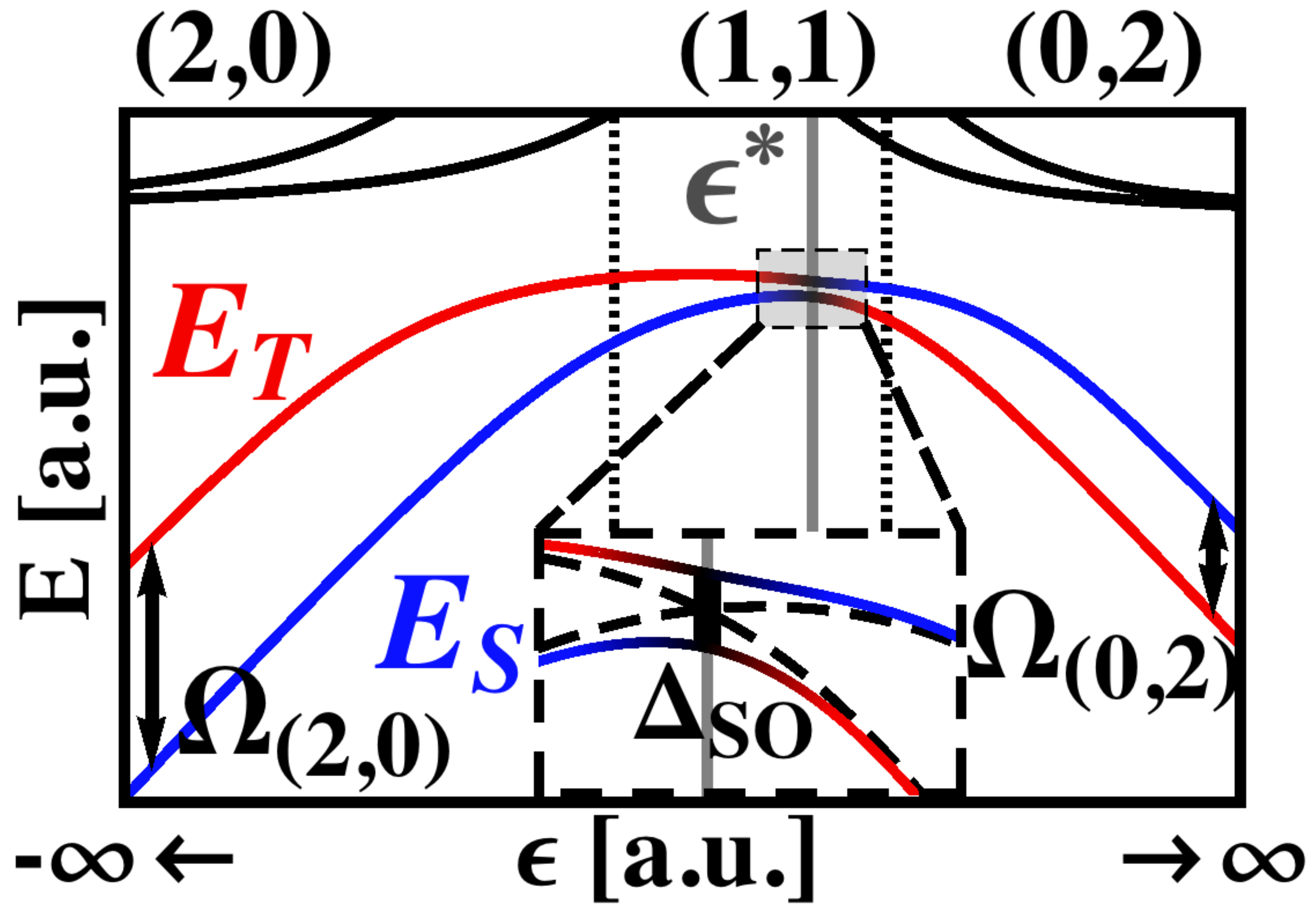}
\caption{Energy diagram of a STQ as a function of the electrostatic bias $\epsilon$ according to \eref{C6-eq:Ham} and \eref{C6-eq:SO}. The blue and red lines describe the energies of the lowest singlet $E_{S}$ and spinless triplet $E_{T}$; black lines show excited states. The left dotted line labels the charge transition point at $\epsilon=-U_L$, where $E_{S_{1,1}}$ and $E_{S_{2,0}}$ have the same energies (similarly $E_{T_{1,1}}$ and $E_{T_{0,2}}$ have equal energies at $\epsilon=U_R$). We obtain a $\left(2,0\right)$ singlet ground state at $\epsilon<0$, while $\epsilon>0$ favors the $\left(0,2\right)$ triplet. $E_S$ and $E_T$ cross at $\epsilon^*$. SOIs couple $E_{S}$ and $E_{T}$. The inset shows the region around $\epsilon^*$.  The dashed curves are energy levels in the absence of  SOIs.
\label{C6-fig:2}}
\end{figure}

\fref{C6-fig:2} shows a typical energy diagram of the $s_z=0$ subspace in the charge configurations $\left(2,0\right)$, $\left(1,1\right)$, and $\left(0,2\right)$. \eref{C6-eq:Ham} describes a state crossing of the singlet $\ket{S}$ and $s_z=0$ triplet state $\ket{T}$. $\ket{S_{2,0}}$ is the ground state deep in $\left(2,0\right)$, while $\ket{T_{0,2}}$ is the ground state in $\left(0,2\right)$. $\ket{S_{1,1}}$ and $\ket{S_{2,0}}$ have the same orbital energies at $\epsilon=-U_L$; $\ket{T_{1,1}}$ and $\ket{T_{0,2}}$ are at equal energies at $\epsilon=U_R$. Similarly, there is a state degeneracy of $\ket{T_{1,1}}$ and $\ket{T_{2,0}}$ at $\epsilon=-\left(U_L+\Omega_{\left(2,0\right)}\right)$. $\ket{S_{1,1}}$ and $\ket{S_{0,2}}$ have the same energy at $\epsilon=U_R+\Omega_{\left(0,2\right)}$. Electron tunneling between the QDs hybridizes states of different charge configurations. The singlet ground state $\ket{S}$ is degenerate with $s_z=0$ triplet state $\ket{T}$ in $\left(1,1\right)$ because the tunnel couplings $t_{s,t}^{L,R}$ are smaller than $U_{L}$, $U_R$, $\Omega_{\left(2,0\right)}$, and $\Omega_{\left(0,2\right)}$. We label this point $\epsilon^{*}$. The next section describes SOIs, which couple $\ket{S}$ and $\ket{T}$ by $\Delta_{\text{so}}$ at $\epsilon^*$.

\section{Calculation of $\Delta_{\text{so}}$
\label{C6-sec:SOI}}
We consider QDs fabricated in the crystal's $\left(0,0,1\right)$ plane. 
The strong confining potential in the z-direction causes interactions between the electron spins and the in-plane momentum components.
SOIs are described by:
\begin{align}
\mathcal{H}_{2}=&\frac{\alpha}{\hbar}\sum_{i=1,2}
	\nonumber
	\left[\sigma_{x^\prime}\wp_{y^\prime}
	-\sigma_{y^\prime} \wp_{x^\prime}\right]_i
	\\&+
	\frac{\beta}{\hbar}\sum_{i=1,2}
	\left[-\sigma_{x^\prime}\wp_{x^\prime}
	+\sigma_{y^\prime} \wp_{y^\prime}\right]_i.
	\label{C6-eq:SO}
\end{align}
The first term, which is called the Rashba SOI,\footnote{
The Rashba constant $\alpha$ and the Dresselhaus constant $\beta$ have the units $\text{Jm}$. $l_{\text{so}}^\alpha=\frac{\hbar^2}{2m\alpha}$ and $l_{\text{so}}^\beta=\frac{\hbar^2}{2m\beta}$ are the Rashba and Dresselhaus spin precession lengths that have the units $\text{m}$.
} is caused by the broken structure inversion symmetry from the confining potential in the z-direction \cite{rashba1960}. The second term, called the Dresselhaus SOI,\cite{Note1}
is present for a crystal lattice without inversion symmetry \cite{dresselhaus1955}. $x^\prime$ and $y^\prime$ label the $\left[1,0,0\right]$-direction and $\left[0,1,0\right]$-direction of the lattice. $\left[1,0,0\right]$ is rotated by the angle $\xi$ from $\mathbf{e}_x$, which is the vector connecting the QD centers (cf. \fref{C6-fig:1}). 
Large spin-orbit (SO) effects are expected when electrons are free to move, which is possible between the QDs in the $\mathbf{e}_{x}$-direction. We consider only the SO contributions that involve the momentum component in the $\mathbf{e}_{x}$-direction ($\wp_x$) and extract from \eref{C6-eq:SO}  
$\widetilde{\mathcal{H}}_2=\frac{\bm{\Xi}}{\hbar}\cdot\sum_{i=1,2}\left[\wp_x\bm{\sigma}\right]_i$, with $\bm{\Xi}=\left(
-\beta\cos\left(2\xi\right),
-\alpha-\beta\cos\left(2\xi\right),0\right)^T$. 
Additional contributions from the in-plane momentum component perpendicular to $\mathbf{e}_x$ are discussed in \aref{C6-app:ADDY}.

$\mathcal{H}_0$ from \eref{C6-eq:H0} dominates over the SO contributions. We apply a unitary transformation $\mathcal{U}=e^{i\left(\mathcal{S}_1+\mathcal{S}_2\right)}$, with
$\mathcal{S}_i=\frac{mx_i}{\hbar^2}\bm{\Xi}\cdot\bm{\sigma}_i$ \cite{khaetskii2000,aleiner2001,levitov2003}. $\mathcal{U}$ was introduced to remove SOIs to second order for confined systems. This transformation turns out to be useful because the transformed Hamiltonian is only position dependent. Note that the equivalent transformation was used in \mrcite{baruffa2010-1}{baruffa2010-2} to show that SOIs act trivially on the $s_z=0$ subspace for a highly symmetric DQD.
The transformed Hamiltonian reads:
\begin{align}
	\label{C6-eq:HTrafo}
	\mathcal{U}\left(\mathcal{H}_0+\mathcal{H}_1+\widetilde{\mathcal{H}}_2\right)\mathcal{U}^\dagger
	=&\mathcal{H}_0-\frac{m}{\hbar^2}\left|\bm{\Xi}\right|^2
	\\&	\nonumber
	+\frac{g\mu_B}{2}
	\sum_{\substack{i=1,2\\j\in\mathbb{N}}}
	\bm{B}_{\text{eff}}^{\left[j\right]}\left(x_i\right)\cdot\bm{\sigma}_i,\\
	\bm{B}_{\text{eff}}^{\left[j\right]}\left(x\right)\equiv
	\frac{1}{j!}\left(\frac{2m}{\hbar^2}x\right)^j&\left[\left(\right.\dots\left(\right.\bm{B}\right.
	\underbrace{\times\left.\bm{\Xi}\right)\times\dots\left.\right)\times\bm{\Xi}}_{j \text{ times}}\left.\right].
	\label{C6-eq:BEFF}
\end{align}
$\mathcal{H}_0$ remains formally unchanged. Besides the constant energy shift $-\frac{m}{\hbar^2}\left|\bm{\Xi}\right|^2$, there are only position dependent terms (note the restriction to the $x$-direction). \eref{C6-eq:HTrafo} couples only states of the same charge sector because the orbital states are strongly confined at the QD's position. We restrict the discussion to the contribution in $\left(1,1\right)$. Contributions from $\left(2,0\right)$ and $\left(0,2\right)$ are negligible, as described in \aref{C6-app:DOC}.  The charge configuration is confined to a small area compared to the SO scale $\left(\frac{\hbar^2}{2m\left|\bm{\Xi}\right|}\right)^2$, with the result that terms in \eref{C6-eq:BEFF} with higher order in $j$ are less important.

The external magnetic field is rotated by the polar angle $\theta$ from the $\left[0,0,1\right]$-direction and the azimuthal angle $\phi$ from $\mathbf{e}_x$ (cf. \fref{C6-fig:1}). We fix the spin quantization axis parallel to $\bm{B}$. The components of \eref{C6-eq:BEFF} that are parallel to the external magnetic field $\left(B_{\text{eff}}^{\left[j\right]}\right)_\shortparallel$ couple $\ket{S_{1,1}}$ and $\ket{T_{1,1}}$, while the perpendicular components couple subspaces of different $s_z$. We assume that the states $\ket{L}$ and $\ket{R}$ are strongly confined at the QD position, with $\Dirac{L}{x}{R}=\Dirac{L}{x^2}{R}=0$, $\Dirac{R}{x}{R}=-\Dirac{L}{x}{L}=a$. We introduce the variances of the orbitals $\Dirac{L}{\left(x-a\right)^2}{L}=\text{var}_L$ and $\Dirac{R}{\left(x+a\right)^2}{R}=\text{var}_R$. Note that the transformation $\mathcal{U}$ in  \eref{C6-eq:HTrafo} modifies also the definitions of the basis states $\ket{L}$ and $\ket{R}$.

The effective Hamiltonian in $\left(1,1\right)$, including SOIs to second order, is written in the basis $\ket{S_{1,1}}$ from \eref{C6-eq:states_begin}, $\ket{T_{1,1}}$ from \eref{C6-eq:state_triplet}, $\ket{T_{1,1}^{+}}=c_{L\uparrow}^\dagger c_{R\uparrow}^\dagger\ket{0}$, and $\ket{T_{1,1}^-}=c_{L\downarrow}^\dagger c_{R\downarrow}^\dagger\ket{0}$:
\begin{widetext}
\begin{align}
\nonumber
\mathcal{H}&_{\left(1,1\right)}=
E_Z
\left(\begin{array}{cc|cc}
0&0&0&0\\
0&0&0&0\\
\hline
0&0&1&0\\
0&0&0&-1
\end{array}\right)
-i\sqrt{2}E_Z \frac{2m a}{\hbar^2}\Xi_\perp
\left(\begin{array}{cc|cc}
0 & 0 & 1 & 1\\
0 & 0 & 0 & 0\\
\hline
-1 & 0 & 0 & 0\\
-1 & 0 & 0 & 0
\end{array}\right)
-E_Z \left(\frac{2m}{\hbar^2}\right)^2\Xi_\perp^2\left(\text{var}_L-\text{var}_R\right)
\left(\begin{array}{cc|cc}
0 & 1 & 0 & 0\\
1 & 0 & 0 & 0\\
\hline
0 & 0 & 1 & 0\\
0 & 0 & 0 & -1
\end{array}\right)
\\\label{C6-eq:HEFFSOI}&
+\sqrt{2}E_Z \left(\frac{2m}{\hbar^2}\right)^2\Xi_\shortparallel\Xi_\perp
\times\left(\begin{array}{cc|cc}
0 & 0 & 
-\frac{\text{var}_L-\text{var}_R}{2} & 
\frac{\text{var}_L-\text{var}_R}{2}\\
0 & 0 & 
\left(\frac{\text{var}_L+\text{var}_R}{2}-2 a^2\right) &  
\left(\frac{\text{var}_L+\text{var}_R}{2}-3 a^2\right)\\\hline
-\frac{\text{var}_L-\text{var}_R}{2} & \left(\frac{\text{var}_L+\text{var}_R}{2}-2 a^2\right) & 0 & 0\\
\frac{\text{var}_-\text{var}_R}{2} & \left(\frac{\text{var}_L+\text{var}_R}{2}-3 a^2\right) & 0 & 0
\end{array}\right),
\end{align}
\end{widetext}
with the Zeeman energy $E_Z=\frac{g\mu_B}{2}\left|\bm{B}\right|$. 
$\Xi_{\shortparallel}=\abs{\bm{\Xi}}\cos\left[\measuredangle\left(\bm{\Xi},\bm{B}\right)\right]$ 
is the component of $\bm{\Xi}$ parallel to $\bm{B}$, and 
$\Xi_{\perp}=\abs{\bm{\Xi}}\sin\left[\measuredangle\left(\bm{\Xi},\bm{B}\right)\right]$ is the component of $\bm{\Xi}$ perpendicular to $\bm{B}$ [all components are determined by the angle $\measuredangle\left(\bm{\Xi},\bm{B}\right)$ between the vectors $\bm{\Xi}$ and $\bm{B}$, cf. \fref{C6-fig:1}].

The first term in \eref{C6-eq:HEFFSOI} represents the Zeeman interaction that shifts $\ket{T_{1,1}^+}$ and $\ket{T_{1,1}^-}$ relative to the $s_z=0$ energy levels. This term dominates over all SO contributions. The second term in \eref{C6-eq:HEFFSOI} couples $\ket{S_{1,1}}$ with $\ket{T_{1,1}^+}$ and $\ket{S_{1,1}}$ with $\ket{T_{1,1}^-}$. This term was discussed in great detail in \mrcite{baruffa2010-1}{baruffa2010-2}. It does not couple $\ket{S_{1,1}}$ and $\ket{T_{1,1}}$. Note that the coupling to the triplet states does not cause an energy shift of $\ket{S_{1,1}}$ in second-order Schrieffer-Wolff perturbation theory \cite{winkler2010} because the couplings between $\ket{S_{1,1}}$ and $\ket{T_{1,1}^+}$ and between $\ket{S_{1,1}}$ and $\ket{T_{1,1}^-}$ cancel each other.

The dominant SO contribution on the $s_z=0$ subspace is obtained from the third term of \eref{C6-eq:HEFFSOI}. This term represents the component of the effective magnetic field parallel to $\bm{B}$, which is second order in the SOI: $\left(B_{\text{eff}}^{\left[2\right]}\right)_{\shortparallel}=-\frac{B}{2}\left(\frac{2m}{\hbar^2}\right)^2\Xi_\perp^2x^2$ [cf. \eref{C6-eq:BEFF}]. $\left(B_{\text{eff}}^{\left[2\right]}\right)_{\shortparallel}$ realizes a direct coupling between $\ket{S_{1,1}}$ and $\ket{T_{1,1}}$: $
\Delta_{\text{so}}\approx E_Z \frac{\text{var}_R-\text{var}_L}{l_{\text{so}}^2}
$. We introduce the length scale $l_{\text{so}}=\frac{\hbar^2}{2m\abs{\Xi_\perp}}$. The fourth term in \eref{C6-eq:HEFFSOI} gives small corrections to $\Delta_{\text{so}}$. \aref{C6-app:ADDY} describes the angular dependency of $\Delta_{\text{so}}$ and extends the analysis of SOIs using all terms of \eref{C6-eq:SO}.

The smallest possible values for $l_{\text{so}}$ are on the order of the Rashba and Dresselhaus spin precession lengths $l_{\text{so}}^{\alpha}$ and $l_{\text{so}}^{\beta}$. Typically, GaAs heterostructures have spin precession lengths $l_{\text{so}}^{\alpha},l_{\text{so}}^{\beta}\gtrsim~1~\mu\text{m}$ for the Rashba and the Dresselhaus SOIs (cf. \aref{C6-app:PARA}). The variances of the orbital wave functions can be approximated using the noninteracting descriptions of electrons that are confined at QDs. The Fock-Darwin states are the solutions of the noninteracting eigenvalue problem of two-dimensional circular QDs \cite{fock1928,darwin1931}. The variances of these wave functions are directly related to the confining potentials as $\text{var}\approx l_0^{2}$, when assuming a harmonic confining potential that has the magnitude $\hbar \omega_0$ with $l_0=\sqrt{\frac{\hbar}{m\omega_0}}\left[1+\left(\frac{e B_z}{2m\omega_0}\right)^2\right]^{-1/4}$. Normal values for strongly confined QDs in GaAs are $\hbar\omega_0=3\ \text{meV}$ and $l_{0}=20~\text{nm}$ \cite{burkard1999}. Weakly confined QDs in GaAs of $\hbar\omega_0=0.1\ \text{meV}$ have $l_{0}=100~\text{nm}$. We obtain, for $l_{\text{so}}=1~\mu\text{m}$ and $B=500~\text{mT}$, $\Delta_{\text{so}}=0.1~\mu\text{eV}$ ($\Delta_{\text{so}}/h\approx 25~\text{MHz}$).

Small-band-gap materials tend to have stronger SOIs. SOIs are, for example, by one order of magnitude larger in InAs than in GaAs ($l_{\text{so}}^\alpha=1.1~\mu\text{m}$ for GaAs and $l_{\text{so}}^\alpha=0.14~\mu\text{m}$ for InAs; cf. \aref{C6-app:PARA}). Furthermore, the variances of the wave functions of InAs QDs are potentially larger than of GaAs QDs due to the smaller effective mass. It should therefore be possible to reach values of $\Delta_{\text{so}}\approx 1~\mu\text{eV}$ ($\Delta_{\text{so}}/h\approx 250~\text{MHz}$).

The coupling between $\ket{S}$ and $\ket{T}$ at $\epsilon^*$ can be approximated by $\Delta_{\text{so}}\approx E_Z\frac{\text{var}_R-\text{var}_L}{l_{\text{so}}^2}$, as one can see from \eref{C6-eq:Ham}. The state coupling is determined by the weights of $\ket{S_{1,1}}$ in $\ket{S}$ and $\ket{T_{1,1}}$ in $\ket{T}$ at $\epsilon^{*}$. $\epsilon^{*}$ is close to the center of $\left(1,1\right)$ because $t_{s,t}^{L,R}$ are much smaller than $U_{L}$, $U_R$, $\Omega_{\left(2,0\right)}$, and $\Omega_{\left(0,2\right)}$. Therefore $\ket{S_{1,1}}$ and $\ket{T_{1,1}}$ have weights close to unity.

In summary, SOIs couple $\ket{S}$ and $\ket{T}$ via their state contributions in $\left(1,1\right)$. There is a second-order coupling through SOIs, describing an effective magnetic field parallel to the external magnetic field $B_{\text{eff}}^\shortparallel$ at the QDs. The magnitude of $B_{\text{eff}}^\shortparallel$ depends on the sizes of the wave functions. $\Delta_{\text{so}}$ is caused by an effective magnetic-field gradient across the DQDs generated from SOIs.

\section{Qubit Manipulations
\label{C6-sec:Manipulate}}
\begin{figure}[htb]
\centering
\includegraphics[width=0.49\textwidth]{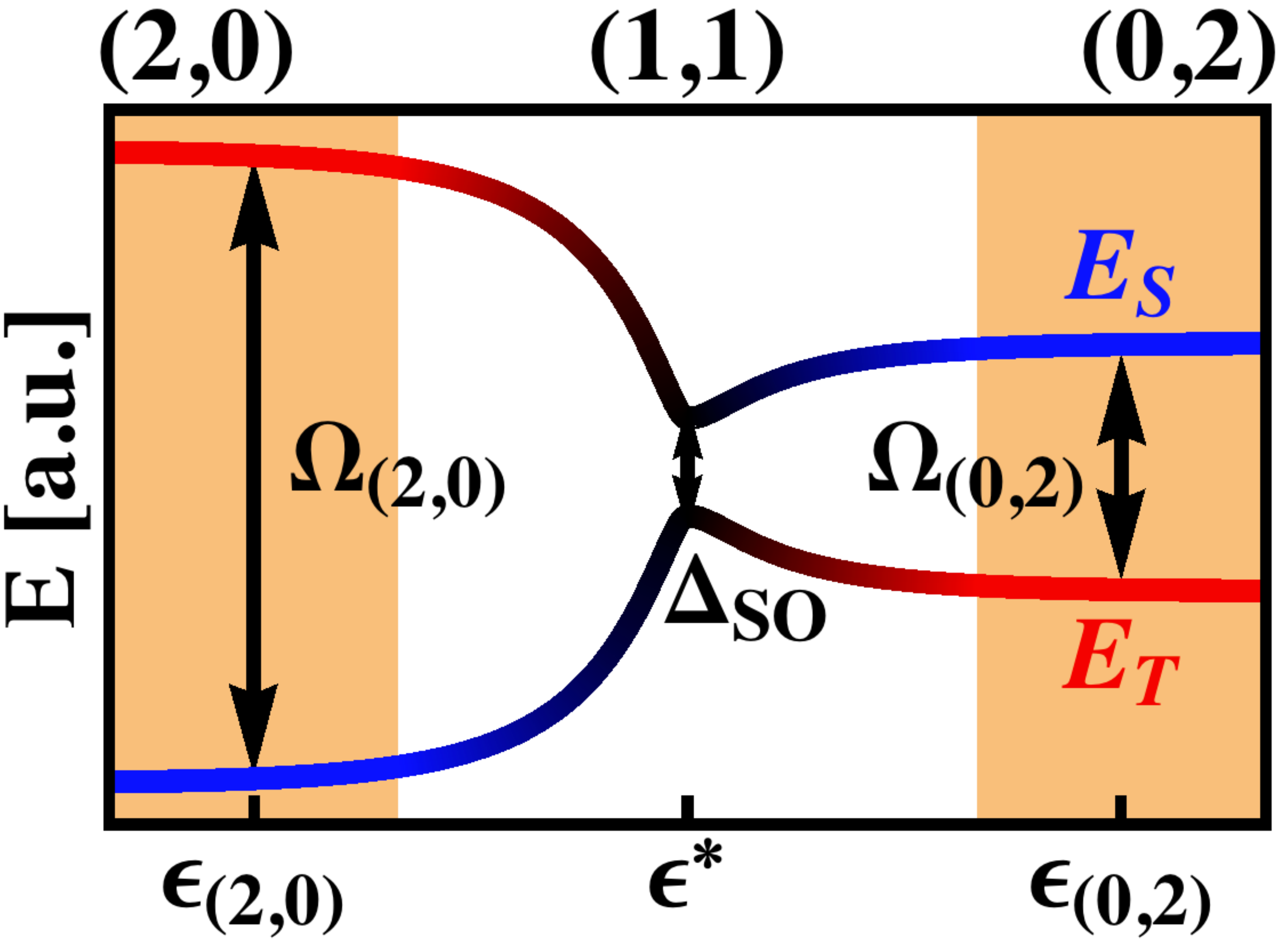}
\caption{Sketch of the energy levels $\ket{S}$ and $\ket{T}$ that encode the ISTQ. The energy levels are shifted compared to \fref{C6-fig:2}, while the energy differences between $E_S$ and $E_T$ remain unchanged at each $\epsilon$. $\ket{S}$ and $\ket{T}$ have equal orbital energies at $\epsilon^*$. SOIs lift the degeneracy and cause an anticrossing $\Delta_{\text{so}}$. $\ket{S}$ is the ground state for $\epsilon<\epsilon^*$, but $\ket{S}$ is the excited state for $\epsilon>\epsilon^*$. We label one point deep in $\left(2,0\right)$ by $\epsilon_{\left(2,0\right)}$ with the energy splitting $\Omega_{\left(2,0\right)}$ [similarly, $\ket{S}$ and $\ket{T}$ have the energy splitting $\Omega_{\left(0,2\right)}$ at $\epsilon_{\left(0,2\right)}$ in $\left(0,2\right)$].
\label{C6-fig:3}}
\end{figure}

An ISTQ encodes a qubit similar to a normal STQ. We identify the singlet state $\ket{S}$ with the logical ``1'' and the $s_z=0$ triplet state $\ket{T}$ with the logical ``0''. Pauli operators are used to describe interactions on the qubit subspace: From this point onward, $\sigma_x=\op{S}{T}+\op{T}{S}$, $\sigma_y=-i\op{S}{T}+i\op{T}{S}$, and $\sigma_z=\op{S}{S}-\op{T}{T}$. A complete set of single-qubit gates together with one maximally entangling two-qubit gate are convenient for universal quantum computation \cite{barenco1995}. \fref{C6-fig:3} shows an energy diagram of the qubit levels as a function of the bias parameter $\epsilon$, which is extracted from \fref{C6-fig:2}. We identify three points that are favorable for qubit manipulations. The qubit states are coupled by a transverse Hamiltonian $\mathcal{H}_{\epsilon^*}=\Delta_{\text{so}}\sigma_x$ at $\epsilon^*$. $\ket{S}$ and $\ket{T}$ are energy eigenstates far from the anticrossing. We label one point in $\left(2,0\right)$ as $\epsilon_{\left(2,0\right)}$ with $\mathcal{H}_{\epsilon_{\left(2,0\right)}}=-\Omega_{\left(2,0\right)}\sigma_z$ [and, similarly, $\epsilon_{\left(0,2\right)}$ in $\left(0,2\right)$ with $\mathcal{H}_{\epsilon_{\left(0,2\right)}}=\Omega_{\left(0,2\right)}\sigma_z$].

\subsection{Single-Qubit Gates}
The ISTQ provides different approaches for single-qubit manipulations. The effective Hamiltonian on the qubit subspace can be tuned using electric gates. Gate manipulations rotate the direction of an effective magnetic field. A magnetic field in the $z$-direction is applied at $\epsilon_{\left(2,0\right)}$ in $\left(2,0\right)$ and $\epsilon_{\left(0,2\right)}$ in $\left(0,2\right)$. $\Omega_{\left(2,0\right)}$ and $\Omega_{\left(0,2\right)}$ correspond to the energy differences of $\ket{S}$ and $\ket{T}$. The effective magnetic-field direction is tilted to the $x$-axis in $\left(1,1\right)$. It points exactly along $\mathbf{e}_x$ at $\epsilon^*$ and has a magnitude $\Delta_{\text{so}}$. Rotations around the $z$-axis and $x$-axis can be generated when the qubit is tuned fast between $\epsilon_{\left(2,0\right)}$, $\epsilon_{\left(0,2\right)}$, and $\epsilon^*$. The qubit manipulation time $\tau$ must be diabatic with the SOI, but adiabatic to the orbital Hamiltonian: 
$h/\Delta_{\text{so}}\gg\tau\gg h/\Omega_{\left(2,0\right)},h/\Omega_{\left(0,2\right)}$ \cite{taylor2007}. The time scale of single-qubit gates is determined by $h/\Omega_{\left(2,0\right)}$, $h/\Omega_{\left(0,2\right)}$, and $h/\Delta_{\text{so}}$; it should be in the range of $10$ MHz to a few GHz. Larger values make the gates too fast to be controlled by electronics. Smaller values require long gate times.

We describe two other possibilities for single-qubit control that are practical if $\Delta_{\text{so}}$ is either very large or very small. A large value of $\Delta_{\text{so}}$ permits resonant Rabi driving, which has already been successful for a qubit encoded in triple QDs \cite{medford2013,taylor2013}. The effective Hamiltonian at $\epsilon^*$ is $\mathcal{H}=\Omega\left(\epsilon\right)\sigma_z+\Delta_{\text{so}} \sigma_x$. Transitions are driven by $\Omega\left(\epsilon\right)=2\Omega_0\cos\left(2\Delta_{\text{so}} t/\hbar+\psi\right)$. If $\Omega_0\ll\Delta_{\text{so}}$, then one obtains after a rotating wave approximation the static Hamiltonian $\mathcal{H}^{\prime}=\Omega_0\left[-\sigma_y\sin\left(\psi\right)+\sigma_z\cos\left(\psi\right)\right]$. A universal set of single-qubit gates can be generated when the phase $\psi$ is adjusted.

Rabi driving becomes impractical for small $\Delta_{\text{so}}$ because the gate times increase. We propose another possibility of driven gates that are described by the Landau-Zener (LZ) model \cite{shevchenko2010,vitanov1996,ribeiro2010}. Traversing the anticrossing in a time similar to $\tau=h/\Delta_{\text{so}}$ generates single-qubit rotations. For large transition amplitudes, as for the sweep from $\epsilon_{\left(2,0\right)}$ to $\epsilon_{\left(0,2\right)}$, the time evolution \cite{shevchenko2010,vitanov1996,ribeiro2010},
\begin{equation}
	\mathcal{U}_{LZ}=e^{-i\zeta_{R}\sigma_z} e^{-i\gamma\sigma_y} e^{-i\zeta_{L}\sigma_z},
	\label{C6-eq:LZ}
\end{equation}
is decomposed into phase accumulations (through $\zeta_R$ and $\zeta_L$) and one rotation around an orthogonal axis. The phase accumulations $\zeta_R$ and $\zeta_L$ are determined by the adiabatic evolution under the energy splitting $\Omega\left(t\right)\sigma_z$ and the St\"uckelberg phase. The essential part is the rotation around the $y$-axis by the angle $\gamma=\gamma_{LZ}+\pi/2$, with $\sin\left(\gamma_{LZ}\right)=\sqrt{P_{LZ}}$, $P_{LZ}=e^{-\frac{2\Delta_{\text{so}}^2}{\hbar v}}$. $v=\left.dE/dt\right|_{\epsilon^{*}}$ is the linearized velocity at $\epsilon^{*}$. For example, the state $\ket{0}$ is transferred to an equal superposition of $\ket{0}$ and $\ket{1}$ for $P_{LZ}=\frac{1}{2}$. 

\subsection{Two-Qubit Gates}
\begin{figure}[htb]
\centering
\includegraphics[width=0.49\textwidth]{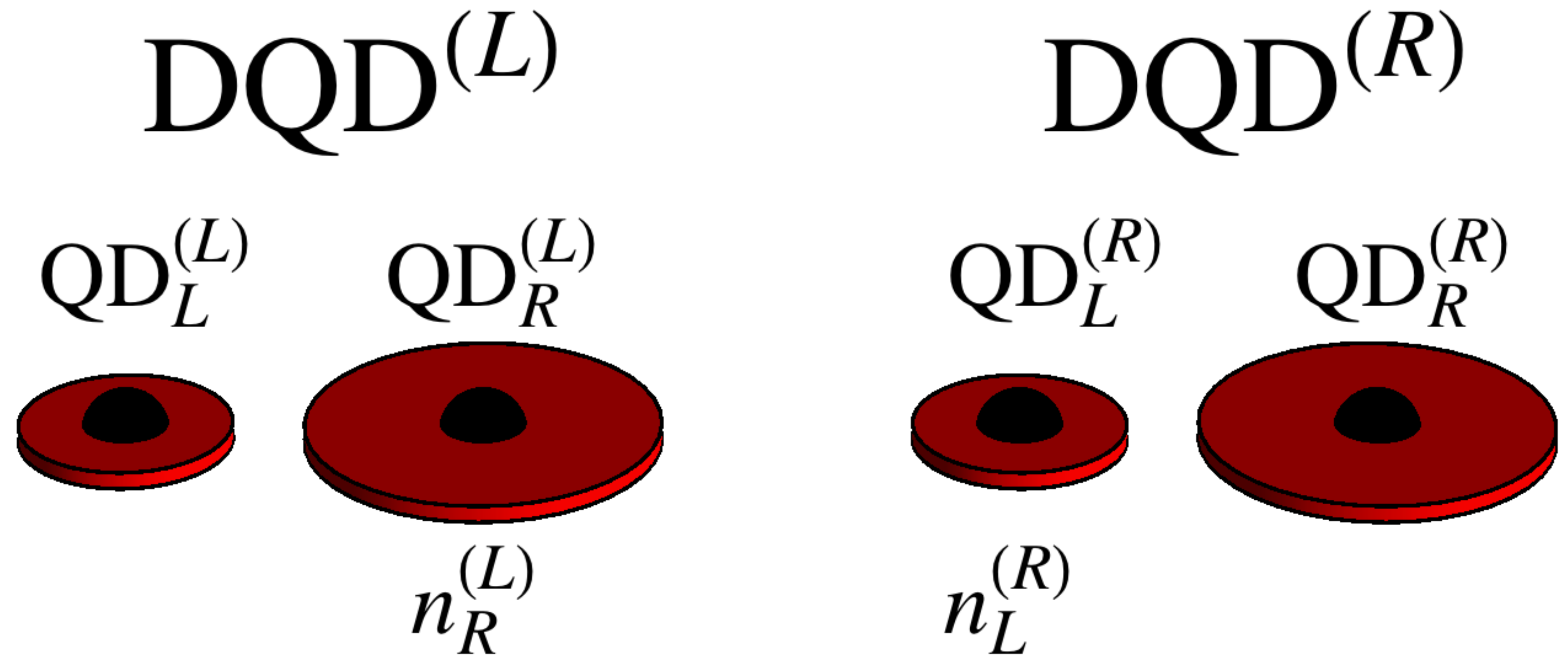}
\caption{Two DQDs [labeled by $\left(L\right)$ and $\left(R\right)$] encode two ISTQs, which are coupled using Coulomb interactions. $\text{QD}_R\pos{L}$ and $\text{QD}_L\pos{R}$ are closest to each other and the electron configurations at these QDs ($n_{R}\pos{L}$ and $n_{L}\pos{R}$) dominate the interaction between the qubits [cf. \eref{C6-eq:CouplEff}].
\label{C6-fig:4}}
\end{figure}

Two-qubit gates can be realized using Coulomb interactions between two ISTQs \cite{taylor2005}. We consider a linear arrangement of four QDs and label the two DQDs by $\left(L\right)$ and $\left(R\right)$ (cf. \fref{C6-fig:4}). $\text{QD}_R\pos{L}$ and $\text{QD}_L\pos{R}$ are closest to each other, and the electron configurations $n_{R}\pos{L}$ at $\text{QD}_R\pos{L}$ and $n_{L}\pos{R}$ at $\text{QD}_L\pos{R}$ dominate the Coulomb coupling between the ISTQs \cite{taylor2013,pal2014}: $
\mathcal{H}_{\text{int}}=
\frac{e^2}{4\pi\epsilon_0\epsilon_r d}
n_{R}\pos{L}n_{L}\pos{R}$. $d$ is the distance between $\text{QD}_R\pos{L}$ and $\text{QD}_L\pos{R}$, $\epsilon_0$ is the dielectric constant, and $\epsilon_r$ is the relative permittivity. $\mathcal{H}_{int}$ leaves the spin at $\text{ISTQ}\pos{L}$ and the spin at $\text{ISTQ}\pos{R}$ unchanged and can only cause the effective interaction
$\mathcal{C}\sigma_z^{\left(1\right)}\sigma_z^{\left(2\right)}$ up to local energy shifts.\footnote{
SOIs mix the spin part and the orbital part of the wave functions, and they also enable effective two-qubit interactions other than $\sigma_z\pos{1}\sigma_z\pos{2}$. We neglect SO contributions for the construction of two-qubit interactions because $\mathcal{H}_2$ from \eref{C6-eq:SO} is weak.
} $\mathcal{C}$ has finite values only when $\ket{S\pos{L}}$ has a charge configuration that differs from that of $\ket{T\pos{L}}$ and $\ket{S\pos{R}}$ has a charge configuration that differs from that of $\ket{T\pos{R}}$ [cf. \eref{C6-eq:Ham}]: \begin{align}
\nonumber
\mathcal{C}=&\frac{e^2}{16\pi\epsilon_0\epsilon_rd}
\left[
\Dirac{S\pos{L}}{n_{R}\pos{L}}{S\pos{L}}-
\Dirac{T\pos{L}}{n_{R}\pos{L}}{T\pos{L}}
\right]\\&\times
\left[
\Dirac{S\pos{R}}{n_{L}\pos{R}}{S\pos{R}}-
\Dirac{T\pos{R}}{n_{L}\pos{R}}{T\pos{R}}
\right]
\label{C6-eq:CouplEff}.
\end{align}
We discuss $\mathcal{C}$, with $\text{STQ}\pos{L}$ and $\text{STQ}\pos{R}$ at $\epsilon^*$, as an example. $\text{QD}_{R}\pos{L}$ has a higher occupation in $\ket{T\pos{L}}$ than in $\ket{S\pos{L}}$ because the doubly occupied triplet in $\left(0,2\right)\pos{L}$ is favored over the doubly occupied singlet. The opposite effect is true for $\text{QD}_{L}\pos{R}$, with a higher electron configuration at $\text{QD}_{L}\pos{R}$ for $\ket{S\pos{R}}$ than for $\ket{T\pos{R}}$. The magnitude of $\mathcal{C}$ strongly depends on the material and the DQD setup. Two electrons with the distance $d=200~\text{nm}$ interact with $\mathcal{C}\approx 100~\mu\text{eV}$ for GaAs and InAs heterostructures ($\epsilon_r=12.9$ for GaAs and $\epsilon_r=15.2$ for InAs \cite{ioffe}). $\mathcal{C}$ is by orders of magnitudes smaller for ISTQs. We assume that $\mathcal{C}/\Delta_{\text{so}}=\frac{1}{10}$ can be reached.

We construct an entangling gate for ISTQs that is similar to common STQs \cite{shulman2012}. Both STQs are pulsed to the transition region of  $\left(1,1\right)$ and $\left(0,2\right)$ with an effective Hamiltonian $\mathcal{H}=\Omega\pos{L}\sigma_z\pos{L}+\Omega\pos{R}\sigma_z\pos{R}+\mathcal{C}\sigma_z\pos{L}\sigma_z\pos{R}$. A $\text{CPHASE}$ gate is generated after the waiting time $t=\frac{h}{8\mathcal{C}}$. This description is valid away from $\epsilon^*$. Directly at $\epsilon^*$, driven entangling operations are permitted through the Hamiltonian $\mathcal{H}=
\Omega\pos{L}\sigma_z\pos{L}+
\Delta_{\text{so}}\pos{L}\sigma_x\pos{L}+
\Omega\pos{R}\sigma_z\pos{R}+
\Delta_{\text{so}}\pos{R}\sigma_x\pos{R}+
\mathcal{C}\sigma_z\pos{L}\sigma_z\pos{R}$. For $\left|\Delta_{\text{so}}\pos{L}-\Delta_{\text{so}}\pos{R}\right|\gg \mathcal{C}$, one possible two-qubit gate is obtained when qubit $\left(L\right)$ is driven with the frequency $2\Delta_{\text{so}}\pos{R}/h$. These driven gates are popular for superconducting qubits \cite{paraoanu2006,rigetti2010,chow2011,chow2012}. The requirement is again that $\Delta_{\text{so}}\pos{L}$ and $\Delta_{\text{so}}\pos{R}$ reach magnitudes of $\mu\text{eV}$ to obtain fast gate operations.

\section{Discussion and Conclusion
\label{C6-sec:Discussion}}
An ISTQ with a finite $\Delta_{\text{so}}$ provides universal control of the $s_z=0$ subspace. Operations mainly at $\epsilon^*$ in $\left(1,1\right)$ and $\epsilon_{\left(0,2\right)}$ in $\left(0,2\right)$ are very favorable because the qubit is protected from small fluctuations in $\epsilon$. $\Omega_{\left(0,2\right)}/h$ in $\left(0,2\right)$ should not exceed a few GHz to control phase accumulations at $\epsilon_{\left(0,2\right)}$. Note that the out-of-plane magnetic-field component $B_z$ determines the magnitude of $\Omega_{\left(0,2\right)}$. Obtaining large $\Delta_{\text{so}}$ is most critical. The size of $\Delta_{\text{so}}$ depends on the confining energies of the QDs and the magnitude of the SOIs. Values of $\Delta_{\text{so}}\approx\mu\text{eV}$ will be needed for driven Rabi gates. We showed that these magnitudes are obtained for two QDs differing strongly in size. This setup is also promising due to other reasons. Strongly confined QDs are ideal for the initialization and readout of STQs. A weakly confined QD can be very useful for qubit manipulations (cf. also \rcite{mehl2013}).

Hyperfine interactions influence qubits in the ISTQ encodings. Nuclear spins couple to the electrons that are confined at QDs by creating local magnetic-field fluctuations $\bm{B}_{\text{hyp}}$. $\bm{B}_{\text{hyp}}$ can be considered as static during one measurement because the nuclear magnetic-field fluctuations are low frequency, but $\bm{B}_{\text{hyp}}$ gives random contributions between successive measurements \cite{coish2005,neder2011}. An approximation for the rms of the component parallel to the external magnetic field that couples to an electron at a QD is $\delta B_{\text{hyp}}^{\shortparallel}\left(\text{QD}\right)= \frac{\sum_{\nu}\mathcal{B}_\nu \sqrt{I_\nu\left(I_\nu+1\right)}}{\sqrt{N}}$ \cite{merkulov2002,taylor2007}. $\nu$ labels the different nuclear spin isotopes of the semiconductor, which have the spin $I$. $\mathcal{B}$ contains material-dependent coupling constants of the isotope, and $N$ is the number of nuclei interacting with an electron that is confined at a QD. If $\text{QD}_L$ and $\text{QD}_R$ have different components of $\bm{B}_{\text{hyp}}$ parallel to the external magnetic field, then for ISTQs the states $\ket{S_{1,1}}$ from \eref{C6-eq:states_begin} and $\ket{T_{1,1}}$ from \eref{C6-eq:state_triplet} are coupled equivalently to $\Delta_{\text{so}}$ by $g\mu_B\left[
B_{\text{hyp}}^{\shortparallel}\left(\text{QD}_L\right)
- B_{\text{hyp}}^{\shortparallel}\left(\text{QD}_R\right)\right]$. In the analysis of many measurements, the rms values of the fluctuations $\delta B_{\text{hyp}}^{\shortparallel}\left(\text{QD}_L\right)$ and $\delta B_{\text{hyp}}^{\shortparallel}\left(\text{QD}_R\right)$ will be detected when assuming independent fluctuations of the magnetic fields at $\text{QD}_L$ and $\text{QD}_R$. We arrive at an effective coupling element between $\ket{S_{1,1}}$ and $\ket{T_{1,1}}$:
\begin{align}
\Delta_{\text{hyp}}=g\mu_B\sqrt{ 
\left[\delta B_{\text{hyp}}^{\shortparallel}\left(\text{QD}_L\right)\right]^2
+ 
\left[\delta B_{\text{hyp}}^{\shortparallel}\left(\text{QD}_R\right)\right]^2}.
\end{align}
An electron at a GaAs QD typically interacts with $10^{6}$ nuclear spins, which gives $
\delta B^\shortparallel_{\text{hyp}}\left(\text{QD}_L\right)=
\delta B^\shortparallel_{\text{hyp}}\left(\text{QD}_R\right)
\approx5~\text{mT}$ and $\Delta_{\text{hyp}}\approx 100~\text{neV}$ for a symmetric DQD.\cite{assali2011}

A weakly confined QD has a smaller uncertainty $\delta B_{hyp}^\shortparallel\left(\text{QD}\right)$ because the electron wave function interacts with more nuclear spins. This size effect will, however, not affect $\Delta_{\text{hyp}}$ significantly for the ISTQ, because the size of only one of the QDs will increase compared to a normal DQD setup, and $\delta B_{hyp}^\shortparallel\left(\text{QD}\right)$ changes only by $N^{-\frac{1}{2}}$. InAs QDs have larger $\Delta_{\text{hyp}}$ than GaAs QDs. Indium isotopes are spin-9/2 nuclei, in contrast to Ga and As nuclei that are spin-3/2. Because of the equivalent influences of hyperfine interactions and SOIs, $\Delta_{\text{so}}$ should be significantly larger than $\Delta_{\text{hyp}}$ to allow high fidelity qubit gates. Our estimates of $\Delta_{\text{so}}=1~\mu\text{eV}$ and $\mathcal{C}/\Delta_{\text{so}}=\frac{1}{10}$ suggest that $\Delta_{\text{hyp}}$ and $\mathcal{C}$ have the same order of magnitude for uncorrected nuclear magnetic fields. Fortunately, many methods are known to reduce the uncertainty of the nuclear magnetic-field distributions by orders of magnitude.\cite{bluhm2010,shulman2014} Additionally, refocusing techniques can be applied to correct for small $\Delta_{\text{hyp}}$ because the magnetic field fluctuations are low frequency \cite{neder2011}.

Charge noise is another source of decoherence. The filling and unfilling of charge traps cause fluctuating electric fields at the positions of the DQDs. If the qubit is operated as a charge qubit, then charge noise dephases the ISTQ \cite{coish2005,hu2006,mehl2013}. Charge fluctuations are dominantly low frequency and lead typically to energy shifts $\delta E_C=\mu eV$ between different charge states \cite{petersson2010,dial2013}. The phase coherences between charge states are lost within a few $\text{ns}$. The most significant influence of charge noise can be described by small fluctuations in $\epsilon$ \cite{dial2013}. Charge noise is less important at $\epsilon^*$, $\epsilon_{\left(2,0\right)}$, and $\epsilon_{\left(0,2\right)}$ because small fluctuations in $\epsilon$ do not dephase the qubit.

In summary, we have discussed a two-electron qubit encoding in the $s_z=0$ subspace for an ISTQ. The out-of-plane magnetic field is used to generate a level crossing of $\ket{S}$ and $\ket{T}$ that is not present for normal STQs. SOIs couple $\ket{S}$ and $\ket{T}$ if the sizes of the QDs differ. Different variances of the wave functions of the QD orbitals cause an effective magnetic-field difference across the DQD. A DQD that consists of two unequal QDs can be a promising spin qubit also for other reasons. It has one QD with a large singlet-triplet splitting and one QD with a small singlet-triplet splitting already without external magnetic fields. The strongly confined QD is ideal for the qubit initialization and the readout, while the weakly confined QD is suitable for the qubit manipulations. We suggest ISTQs in GaAs and InAs because they provide sufficiently large $\Delta_{\text{so}}$.

Hyperfine interactions and charge noise dephase ISTQs. Hyperfine interactions cause dephasing mainly in $\left(1,1\right)$ through low-frequency magnetic-field fluctuations. Nuclear spins and SOIs couple to ISTQs in the same way. It is very important to fabricate ISTQs, where $\Delta_{\text{so}}$ is larger than the fluctuation $\Delta_{\text{hyp}}$ from nuclear spins. Nuclear spin noise can be refocused for ISTQs because fluctuations in $\Delta_{\text{hyp}}$ are low frequency. Charge noise dephases the qubit in the transition region between different charge sectors. Charge noise will be dealt with most efficiently if the ISTQ is operated only at $\epsilon^*$ and deep in $\left(0,2\right)$. All qubit operations require fast manipulation periods between different charge configurations, which has been achieved in previous experiments \cite{shi2014,kim2014}. Motivated by the search for alternative spin qubit designs \cite{shi2014,kim2014,higginbotham2014}, we are hopeful that DQDs are explored where the QDs differ significantly in size. Realizing an ISTQ in a DQD of two different QDs will be possible by simply tilting the magnetic field out of plane. The perspective of universal electrostatic control which uses only a static SO-induced anticrossing should further motivate the exploration of this setup.

\textit{Acknowledgments} \textthreequartersemdash\ 
We are grateful for support from the Alexander von Humboldt foundation.

\begin{appendix}
\section{Full Calculation of $\Delta_{\text{so}}$ from SOIs
\label{C6-app:ADDY}}

This section extends the calculation of $\Delta_{\text{so}}$ from to the main text. Here we take into account that the DQD system is not only one dimensional. Besides 
$\widetilde{\mathcal{H}}_2=\frac{\bm{\Xi}}{\hbar}\cdot\sum_{i=1,2}\left[\wp_x\bm{\sigma}\right]_i$, with 
$\bm{\Xi}=
\left(-\beta\cos\left(2\xi\right),-\alpha-\beta\sin\left(2\xi\right),0\right)^T$, describing the momentum component connecting the QDs, there is also the in-plane perpendicular momentum component 
$\widetilde{\widetilde{\mathcal{H}}}_2=\frac{\bm{\Psi}}{\hbar}\cdot\sum_{i=1,2}\left[\wp_y\bm{\sigma}\right]_i$, with 
$\bm{\Psi}=
\big(\alpha-\beta\sin\left(2\xi\right),\beta\cos\left(2\xi\right),0\big)^T$. $\widetilde{\widetilde{\mathcal{H}}}_2$ matters for QDs, in which the electrons have space to move in the $y$-direction. Now, we discuss the extreme case of circular QDs. We assume, additionally to the properties of $\ket{L}$ and $\ket{R}$ that were introduced in \sref{C6-sec:SOI}, $\Dirac{L}{y}{L}=\Dirac{R}{y}{R}=0$, $\Dirac{L}{y^2}{L}=\text{var}_L$, $\Dirac{R}{y^2}{R}=\text{var}_R$, and that $\ket{L}$ and $\ket{R}$ are separable into an x-part and y-part.

We apply the transformation 
$\mathcal{U}=e^{i\left(\mathcal{S}_1+\mathcal{S}_2\right)}$, with $\mathcal{S}_i=\frac{m}{\hbar^2}\left[\wp_x\bm{\Xi}+\wp_y\bm{\Psi}\right]_i\cdot\bm{\sigma}_i$. The transformed Hamiltonian $\mathcal{U}\left(\mathcal{H}_0+\mathcal{H}_1+\widetilde{\mathcal{H}}_2+\widetilde{\widetilde{\mathcal{H}}}_2\right)\mathcal{U}^\dagger$ contains similar terms as in \eref{C6-eq:HTrafo}. Formally, $\mathcal{H}_0$ remains unchanged, and there is an overal energy shift $-\frac{m}{\hbar^2}\left(\left|\bm{\Xi}\right|^2+\left|\bm{\Psi}\right|^2\right)$. $\mathcal{H}_1$ from \eref{C6-eq:Zeem} gives a position-dependent magnetic field,
\begin{widetext}
\begin{align}
\mathcal{U}\mathcal{H}_1\mathcal{U}^\dagger=&
\frac{g\mu_B}{2}
\sum_{\substack{i=1,2\\j\in\mathbb{N}}}
\bm{B}_{\text{eff}}^{\left[j\right]}\left(\bm{x}_i\right)
\cdot\bm{\sigma}_i,\\\label{C6-eq:BeffTOT}
\bm{B}_{\text{eff}}^{\left[j\right]}\left(\bm{x}\right)\equiv&
\frac{1}{j!}\left(\frac{2m}{\hbar^2}\right)^j
\left[\left(\dots\left(\bm{B}\right.\right.
\underbrace{\left.\left.\times\left(\bm{\Xi}x+\bm{\Psi}y\right)\right)\dots\right)\times\left(\bm{\Xi}x+\bm{\Psi}y\right)}_{j \text{ times}}\right].
\end{align}
\end{widetext}

We extract from \eref{C6-eq:BeffTOT} the effective magnetic-field component parallel to $\bm{B}$ in second order of the SOIs: 
\begin{align}
\left(B_{\text{eff}}^{\left[2\right]}\right)_\shortparallel\approx-\frac{B}{2}\left(\frac{2m}{\hbar^2}\right)^2
\left(\Xi_\perp^2 x^2+\Psi_\perp^2 y^2\right).
\label{C6-eq:cont1}
\end{align}
\eref{C6-eq:cont1} neglects mixed terms in the position operators ($\sim xy$) and couples $\ket{S_{1,1}}$ and $\ket{T_{1,1}}$ by $\Delta_{\text{so}}^{Z}=E_Z\frac{\text{var}_R-\text{var}_L}{\left(l_{\text{so}}^Z\right)^2}$ with $l_{\text{so}}^Z=\frac{\hbar^2}{2m\sqrt{\Xi_\perp^2+\Psi_\perp^2}}$ (which we call the Zeeman spin precession length). 
$\Xi_\perp=\abs{\bm{\Xi}}\sin\left[\measuredangle\left(\bm{\Xi},\bm{B}\right)\right]$ and 
$\Psi_\perp=\abs{\bm{\Psi}}\sin\left[\measuredangle\left(\bm{\Psi},\bm{B}\right)\right]$ are the components of $\bm{\Xi}$ and $\bm{\Psi}$ perpendicular to the external magnetic field (cf. \fref{C6-fig:1}).
Note that $l_{\text{so}}^Z$ is on the order of the Rashba and Dresselhaus spin precession length, which is smaller than the confining radius of the QD wave functions.

The transformation of $\widetilde{\mathcal{H}}_2+\widetilde{\widetilde{\mathcal{H}}}_2$ adds additional contributions, dominated by:
\begin{align}
\label{C6-eq:helppp}
\frac{m\bm{\Xi}\times\bm{\Psi}}{\hbar^3}\cdot\sum_{i=1,2}\bm{\sigma}_i\left[\left(l_z\right)_i-\frac{m\omega_c}{2}
\left(x^2_i+y^2_i\right)\right],
\end{align} 
with $l_z=p_xy-p_yx$, and $\omega_c=\frac{eB_z}{m}$. Especially the second term in \eref{C6-eq:helppp} couples $\ket{S_{1,1}}$ and $\ket{T_{1,1}}$ directly by an effective magnetic field parallel to $\bm{B}$:
\begin{align}
\left(B_{\text{eff},o}^{\left[2\right]}\right)_\shortparallel=
-\frac{\hbar\omega_c/8}{\frac{g\mu_B}{2}}
\left(\frac{2m}{\hbar^2}\right)^2
\left(\bm{\Xi}\times\bm{\Psi}\right)_\shortparallel
\left(x^2+y^2\right).
\label{C6-eq:cont2}
\end{align}
$\left(\bm{\Xi}\times\bm{\Psi}\right)_\shortparallel$ is the component parallel to $\bm{B}$, which can be positive or negative. $\left(B_{\text{eff},o}^{\left[2\right]}\right)_\shortparallel$ is determined by the orbital contribution of the magnetic field $\hbar\omega_c$ instead of the Zeeman energy $E_Z=\frac{g\mu_B}{2}\left|\bm{B}\right|$. It describes the magnetic field produced by the orbital motion of electrons. We introduce the orbital spin precession length $\left(l_{\text{so}}^o\right)^2=\left(\frac{\hbar^2}{2m}\right)^2\frac{1}{\left(\bm{\Xi}\times\bm{\Psi}\right)_\shortparallel}$, with which we write $\Delta_{\text{so}}^{o}=
\frac{\hbar\omega_c}{2}
\frac{\text{var}_L-\text{var}_R}{\left(l_{\text{so}}^o\right)^2}$.

\eref{C6-eq:cont1} and \eref{C6-eq:cont2} couple $\ket{S_{1,1}}$ and $\ket{T_{1,1}}$ by a magnetic-field gradient across the DQD, similarly to the consideration in the main text:
\begin{align}
\Delta_{\text{so}}=\left(\text{var}_R-\text{var}_L\right)\left(\frac{E_Z}{\left(l_{\text{so}}^Z\right)^2}+\frac{\hbar\omega_c/2}{\left(l_{\text{so}}^o\right)^{2}}\right).
\label{C6-eq:SOFULL}
\end{align}
Whether the Zeeman contribution $\frac{E_Z}{\left(l_{\text{so}}^Z\right)^2}$ or the orbital contribution $\frac{\hbar\omega_c/2}{\left(l_{\text{so}}^o\right)^2}$ dominates \eref{C6-eq:SOFULL} depends in detail on the DQD. The orbital contribution should be dominant if the QDs are circular because $\hbar \omega_c$ is usually larger than $E_Z$: $\left|\frac{\hbar\omega_c}{E_Z}\right|\approx135$ for GaAs and $\left|\frac{\hbar\omega_c}{E_Z}\right|\approx12$ for InAs (cf. \aref{C6-app:PARA}). If the DQD setup prefers one spatial direction, then the Zeeman contribution dominates.

We analyze the angular dependencies of $\Delta_{\text{so}}$, which are influenced by the direction of the magnetic field $\bm{B}$, the orientation of the crystal lattice, and the dot connection axis $\bm{e}_x$ (cf. \fref{C6-fig:1}). The Zeeman spin precession length gives $\left(l_{\text{so}}^{Z}\right)^{-2}\propto \Xi_\perp^2+\Psi_\perp^2=
2\left(\alpha^2+\beta^2\right)
+\sin^2\left(\theta\right)\left\{\alpha^2+\beta^2+2\alpha\beta\sin\left[2\left(\phi-\xi\right)\right]\right\}$. SO contributions are maximal for out-of-plane magnetic fields, but they can vanish for in-plane magnetic fields. This is exactly the case if there is no coordinate of the SO field perpendicular to the magnetic field.
$\left(l_{\text{so}}^{o}\right)^{-2}\propto 
\left(\bm{\Xi}\times\bm{\Psi}\right)_\shortparallel=
\left(\alpha^2-\beta^2\right)\cos\left(\theta\right)$ is independent of the orientation of the crystal lattice. Orbital effects are maximal for out-of-plane magnetic fields, but they vanish for in-plane orientations.

\section{Doubly Occupied Single QDs
\label{C6-app:DOC}}

This section describes the influence of the SOIs in the $\left(2,0\right)$ and $\left(0,2\right)$ configurations when one QD is doubly occupied. A doubly occupied single QD with the center at $\left(a,0,0\right)^T$ is described by
\begin{align}
	\underbrace{\sum_{i=1,2}\left[\frac{\bm{\wp}^2_i}{2m}+V\left(\bm{x}_i\right)\right]+V\left(\bm{x}_1,\bm{x}_2\right)}_{\mathcal{H}_0}+
	\underbrace{\frac{g\mu_B}{2}\bm{B}\cdot\sum_{i=1,2}\bm{\sigma}_i}_{\mathcal{H}_{1}},
\end{align}
and $\widetilde{\mathcal{H}}_2=\frac{\bm{\Xi}}{\hbar}\cdot\sum_{i=1,2}\left[\wp_x\bm{\sigma}\right]_i$. We apply a unitary transformation $\mathcal{U}=e^{i\left(\mathcal{S}_1+\mathcal{S}_2\right)}$, with $\mathcal{S}_i=\frac{m}{\hbar^2}\left(x_i-a\right)\bm{\Xi}\cdot\bm{\sigma}_i$. $-\frac{ma}{\hbar^2}\bm{\Xi}\cdot\bm{\sigma}_i$ generates a constant, position-dependent phase shift of the transformed states. The transformed Hamiltonian,
\begin{align}
\nonumber
\mathcal{U}\left(\mathcal{H}_0+\mathcal{H}_1+\mathcal{H}_2\right)\mathcal{U}^\dagger=&
\mathcal{H}_0
-\frac{m}{\hbar^2}\left|\bm{\Xi}\right|^2
\\&+\frac{g\mu_B}{2}
\sum_{i=1,2}
\bm{B}_{\text{eff}}\left(x_i\right)\cdot\bm{\sigma}_i,
\end{align}
describes a position-dependent magnetic field:
\begin{widetext}
\begin{align}
\bm{B}_{\text{eff}}\left(x\right)\equiv&
\left\{
\left(\begin{array}{c}
\sin\left(2\rho\right)\left\{\frac{1}{2}-\cos\left[\frac{2m\Xi}{\hbar^2}\left(x-a\right)\right]\right\}\\
\sin\left(\rho\right)\sin\left[\frac{2m\Xi}{\hbar^2}\left(x-a\right)\right]\\
0
\end{array}\right)
\right.
\left.
+\left(\begin{array}{c}0\\0\\1\end{array}\right)
\left[\sin^{2}\left(\rho\right)\cos\left(\frac{2m\Xi}{\hbar^2}\left(x-a\right)\right)+\cos^2\left(\rho\right)\right]
\right\},
\end{align}
\end{widetext}
where $\rho$ is the rotation angle between $\bm{B}$ and $\bm{\Xi}$, $\Xi=\abs{\bm{\Xi}}$. Note that there is a simple geometric relation between the angle $\rho$ and the angles $\theta$, $\phi$, and $\xi$ (cf. \fref{C6-fig:1})
$\bm{B}_{\text{eff}}\left(x\right)$ does not couple $\ket{S}$ and $\ket{T}$ below the quadratic order in the position. Here, a different spread of the singlet and triplet wave functions will be seen. We can neglect these contributions to $\Delta_{\text{so}}$ because $\ket{S}$ and $\ket{T}$ have low weights in $\left(2,0\right)$ and $\left(0,2\right)$ at $\epsilon^*$.

\section{Spin-Orbit Parameters
\label{C6-app:PARA}}

This section describes SOIs for typical semiconductor materials to build QDs. We introduce the Rashba and Dresselhaus SOIs for following \mmrcite{winkler2010}{ihn2010}{zwanenburg2013}.
Rashba SOI is caused by the broken structure inversion symmetry through the confining potential. The Rashba parameter $\alpha$ is determined by the confining electric field $\mathcal{E}_z$ and a material constant $\alpha_R$: $\alpha=\alpha_R \mathcal{E}_z$ \cite{ihn2010}. Typical values for $\mathcal{E}_z$ are $0.1~\text{mV nm}^{-1}$. We introduce the Rashba spin precession length $l_{\text{so}}^{\alpha}=\frac{\hbar^2}{2m\alpha}$. Dresselhaus SOI is present for a semiconducting lattice without inversion symmetry. The Dresselhaus parameter $\beta$ is determined by a band parameter $\beta_D$ and the size of the wave function in the z-direction $\left\langle k_z^2\right\rangle$: $\beta=\beta_D\left\langle k_z^2\right\rangle$. Typical values are $\left\langle k_z^2\right\rangle=\left(10~\text{nm}\right)^{-1}$. We introduce the Dresselhaus spin precession length $l_{\text{so}}^{\beta}=\frac{\hbar^2}{2m\beta}$.

Typical parameters for GaAs, Si, and InAs are summarized in \tref{C6-tab:1}. 
Conduction band electrons in Si have weak SOIs. Electrons in GaAs heterostructures have micrometer spin precession lengths. SOIs are by one order of magnitude larger in InAs than in GaAs because InAs has a much
smaller band gap.

\begin{table}[htb]
\centering
\begin{tabular}{l|ccccccc}
\hline
\hline
& g &$m/m_e$&
$\alpha\left[\text{meV nm}\right]$&
$l_{\text{so}}^\alpha\left[\mu\text{m}\right]$&
$\beta\left[\text{meV nm}\right]$&
$l_{\text{so}}^\beta\left[\mu\text{m}\right]$\\
\hline
GaAs & $-0.44$ & $0.067$ & $0.52$ & $1.1$ & $0.28$ & $2.0$\\
Si & $2$ & $0.19$ & $0.01$ & $20$ & - & - \\
InAs & $-14.9$ & $0.023$ & $11.7$ & $0.14$ & $0.27$ & $6.1$\\
\hline
\hline
\end{tabular}
\caption{Parameters for the Rashba ($\alpha$) and the Dresselhaus ($\beta$) SOIs, as described in the main text. The effective mass for the conduction band electron $m$ (compared to the free electron mass $m_e$) and the g-factor are taken from \mmrcite{ihn2010}{zwanenburg2013}{thalakulam2010}. 
The following band parameters are used: 
$\alpha_{R}^{\text{GaAs}}=5.2~\text{e\AA}^2$, 
$\alpha_{R}^{\text{Si}}=0.11~\text{e\AA}^2$, 
$\alpha_{R}^{\text{InAs}}=117.1~\text{e\AA}^2$,
$\beta_{R}^{\text{GaAs}}=5.2~\text{e\AA}^2$, 
and
$\beta_{R}^{\text{InAs}}=117.1~\text{e\AA}^2$ \cite{winkler2010,ihn2010,tahan2005}. We introduce the Rasba spin precession length $l_{\text{so}}^{\alpha}=\frac{\hbar^2}{2m\alpha}$ and the Dresselhaus spin precession length $l_{\text{so}}^{\beta}=\frac{\hbar^2}{2m\beta}$. Si crystals have a center of inversion, which excludes the Dresselhaus SOI.
\label{C6-tab:1}}
\end{table}
\end{appendix}

\bibliography{library}
\end{document}